\documentclass[conference]{IEEEtran}
\usepackage{graphicx}
\usepackage{epsfig}
\ifCLASSINFOpdf
\else
\fi
\usepackage[cmex10]{amsmath}
\newtheorem{theorem}{Theorem}
\usepackage{algorithm}
\usepackage{algorithmic}

\usepackage{mdwmath}
\begin{document}
%
\title{Light-Hierarchy:  The Optimal Structure\\ for Multicast Routing in WDM Mesh Networks
}


\author{\IEEEauthorblockN{Fen Zhou, and Mikl\a'os Moln\a'ar}
\IEEEauthorblockA{IRISA / INSA Rennes\\ Campus de Beaulieu\\
Rennes, France, 35042\\
Email: \{fen.zhou, molnar\}@irisa.fr}
\and
\IEEEauthorblockN{Bernard Cousin}
\IEEEauthorblockA{IRISA / University of Rennes 1\\ Campus de Beaulieu\\
Rennes, France, 35042\\
Email: bernard.cousin@irisa.fr}
}

\maketitle

\begin{abstract}
Based on the false assumption that multicast incapable (MI) nodes could not be traversed twice on the same wavelength, the light-tree structure was always thought to be optimal for multicast routing in sparse splitting Wavelength Division Multiplexing (WDM) networks. In fact, for establishing a multicast session, an MI node could be crosswise visited more than once to switch a light signal towards several destinations with only one wavelength through different input and output pairs. This is called Cross Pair Switching ($CPS$). Thus, a new multicast routing structure \emph{light-hierarchy} is proposed for all-optical multicast routing, which permits the cycles introduced by the $CPS$ capability of MI nodes. We proved that the optimal structure for minimizing the cost of multicast routing is a set of \emph{light-hierarchies} rather than the light-trees in sparse splitting WDM networks. Integer linear programming (ILP) formulations are developed to search the optimal light-hierarchies. Numerical results verified that the light-hierarchy structure could save more cost than the light-tree structure. \\
\end{abstract}


%
\IEEEpeerreviewmaketitle

\section{Introduction}
\label{introdcution}
Optical wavelength division multiplexing (WDM) networking has been identified as an effective technique for future wide area network environments, due to its potential ability to meet rising demands of high bandwidth and low latency communication \cite{bWen2001}. For bandwidth-driven and time sensitive applications such as video-conference, shared workspace, distributed interactive simulation and software upgrading in WDM networks, multicasting is advised~\cite{mRamalho2000}. The purpose of multicast routing is to provide efficient communication services for applications that necessitate the simultaneous transmission of information from one source to multiple destinations, i.e. one-to-many communication~\cite{jyH2002}. To support multicast in WDM networks, the network nodes should be equipped with optical power splitters, which is capable of splitting the incoming light signal into all the outgoing ports. Thus they are called multicast capable nodes (MC)~\cite{rMalli1998}. However, if a light signal is split into $m$ copies, the signal power of one copy will be reduced to or less than $1/m$ of the original signal power~\cite{mAli2000}. Power loss, complicated architecture plus expensive fabrication prevent the availability of splitters on all network nodes. Hence, the drawbacks of light splitters yield to the employment of TaC (Tap and Continue)~\cite{mAli2000Cost} functionality on all the network nodes, which is able to tap into the light signal for local consumption and forward it to only one output port. The nodes only equipped with TaC capacity are called multicast incapable nodes (MI)~\cite{rMalli1998}.

In full splitting WDM networks, where all nodes can split, light-tree is the optimal structure for establishing a multicast session. As proved in~\cite{xhJia2001}, it is a Steiner problem and NP-hard to find the light-tree with the optimal wavelength channel cost. In spare splitting WDM networks~\cite{rMalli1998}, only a small ratio of nodes are MC nodes while the rest are MI nodes. In this case, the light-tree~\cite{LHSahasrabuddhe1999} structure was still thought to be optimal for all-optical multicast routing based on the false assumption that the MI nodes could not be traversed twice on the same wavelength. Thus, a set of light-trees (i.e. light-forest~\cite{xjzhang2000}) are proposed to solve the multicast routing problem. For instance, ILP formulations are proposed in~\cite{yOliver2005} to compute the loss balanced light-trees for multicast routing with multi-drops constraints. In~\cite{mtChen2008} ILP formulations are developed to search the optimal light-trees solution for multicast routing under delay constraints. However paper~\cite{fZhou2009} pointed out that the light-tree structure is not optimal if there are high degree (no less than 4) MI nodes in the network. And papers~\cite{mAli2000Cost,yLi2008Globecom} proposed the light-trial for multicasting in WDM networks without splitters. However, the so called light-trail structure does not consider and make use of the advantages of light splitters. Thus it is not efficient for multicast in the case of sparse splitting configuration. 

In this paper, a new all-optical multicast routing structure named light-hierarchy is introduced for cost-optimal multicast in sparse splitting WDM networks. Costly wavelength converters are not available in our optical networks. Besides, the fiber cable between two neighbor nodes consists of two oppositely directed fiber links. Contrary to the traditional assumption, an MI node can be viewed as a special branching node, because it could be crosswise visited more than once to forward a light signal towards several multicast members with only one wavelength by employing different input and output ports pairs. This is called Cross Pair Switching. As a result, the cost optimal multicast structure in sparse splitting WDM networks is no longer light-trees, but a set of light-hierarchies, where cycles may exist. This is the basic motivation of our study. ILP formulations are developed for computing the cost optimal light-hierarchies.

 The rest of the paper is organized as follows. First, the multicast routing problem in WDM networks is formulated in Section~\ref{sec: All-Optical Multicast Routing With Sparse Light Splitting}. Then the Cross Pair Switching of MI nodes is introduced in Section~\ref{sec: Light-Hierarchy: A New Structure For All-Optical Multicast Routing}, which leads to a new multicast structure: light-hierarchy. In Section~\ref{sec: ILP Formulation of Light-Hierarchy}, ILP formulations are developed to search the optimal light-hierarchy. Simulations are done in Section~\ref{sec: Simulation and Performance Evaluation} to compare light-hierarchy and light-tree. Finally, the paper is concluded in Section~\ref{sec: Conclusion}.

\section{All-Optical Multicast Routing With Sparse Light Splitting}
\label{sec: All-Optical Multicast Routing With Sparse Light Splitting}
A spare light splitting WDM network can be modeled by an undirected graph $G (V, E, c, W)$. $V$ represents the vertex-set of $G$. Each node $v \in V$ is either an MI or an MC node.
\begin{eqnarray}
    \nonumber V = \{v | v~\textrm{is}~MI~or~v~\textrm{is}~MC\}\\
        |V|=N, |E|=M
\end{eqnarray}
$E$ represents the edge-set of $G$, which corresponds to the fiber links between the nodes in the network. Two directed optical fibers are configured for opposite directions in each link. $W$ denotes the set of wavelengths supported in each fiber link. Each edge $e \in E$ is associated with a cost functions $c(e)$.
We consider a multicast session $ms(s, D)$, which requests for setting up a set of multicast distribution light-structures (e.g., light-trees) from the source $s$ to a group of destinations $D$ simultaneously under the following optical constraints:
(i) Wavelength Continuity Constraint. In the absence of wavelength converters, the same wavelength should be used continuously on all the links of a light-structure.
(ii) Distinct Wavelength Constraint. Two light-trees should be assigned with different wavelengths unless there are edge disjoint.
(iii) Sparse Light Splitting Constraint.
Without loss of generality, assume $k$ light-structures $LS_{i}(s,D_{i})$ are computed in sequence for $ms(s,D)$, where $i \in [1,k]$, and $1 \leq k \leq |D|$. Regarding the optimization of network resources, the total cost (i.e., the wavelength channel cost consumed per multicast session) should be minimized. The total cost can be calculated by the cost sum of all the light-structures built for $ms(s,D)$.
 \begin{eqnarray}
 \label{equation: cost}
 \nonumber c\big(ms(s,D)\big) &=& \sum_{i=1}^{k}{c(LS_{i})}\\
                    &=& \sum_{i=1}^{k}\sum_{e\in LS_{i}}{c(e)}
 \end{eqnarray}

\section{Light-Hierarchy: A New Structure For All-Optical Multicast Routing}
\label{sec: Light-Hierarchy: A New Structure For All-Optical Multicast Routing}
An MI node is only equipped with TaC capacity and thus it is incapable of light splitting. In absence of wavelength converters, the same wavelength should be retained along all the links in a light-tree. Therefore, the MI nodes were thought to be able to only act either as a leaf node or as a two degree intermediate node in a light-tree. Nevertheless, it is very interesting to find that an MI node can work as a special branching node by using Cross Pair Switching.

\subsection{Cross Pair Switching}
\label{subsec: Cross Pair Switching}

\begin{figure}
\centering
\includegraphics[width=.3\textwidth]{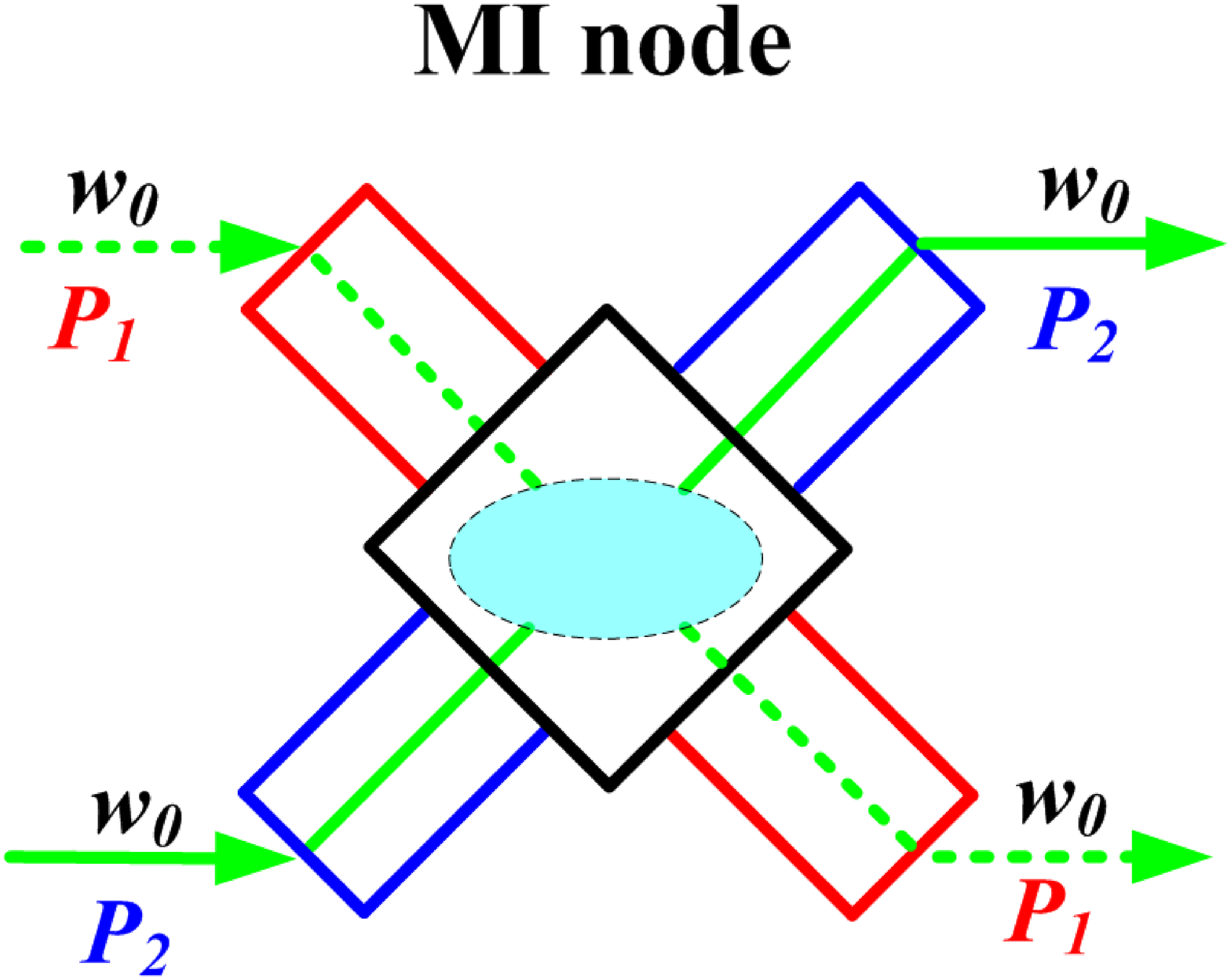}
\caption{Cross Pair Switching of an MI node}
\label{fig: cross-switching}
\end{figure}

Since two oppositely directed optical fibers are placed between each two neighbor nodes in WDM networks, a non-terminal MI node is connected with at least two incoming links as well as two outgoing links as shown in Fig.~\ref{fig: cross-switching}. Assume two signals on the same wavelength $w_{0}$ come from two different lightpaths $P_{1}$ and $P_{2}$. They enter two different input ports of an MI node. As we can see in Fig.~\ref{fig: cross-switching}, with the help of vacant port pairs, the MI node is able to switch these two signals into two outgoing ports without any conflict. Note that the signals are still on the same wavelength $w_{0}$, but forwarded to different successor nodes. Here, we call it as Cross Pair Switching ($CPS$). Based on the $CPS$ capacity of MI nodes, an MI node could connect two successor nodes in a light-structure (the same wavelength should be respected along all the paths in a light-structure) by making use of different input and output port pairs. In this case, an MI node can be traversed twice, then the multicast structure will be no longer a light-tree, but a light-hierarchy, where cycles may exist.

\begin{figure}
\centering
\includegraphics[width=.4\textwidth]{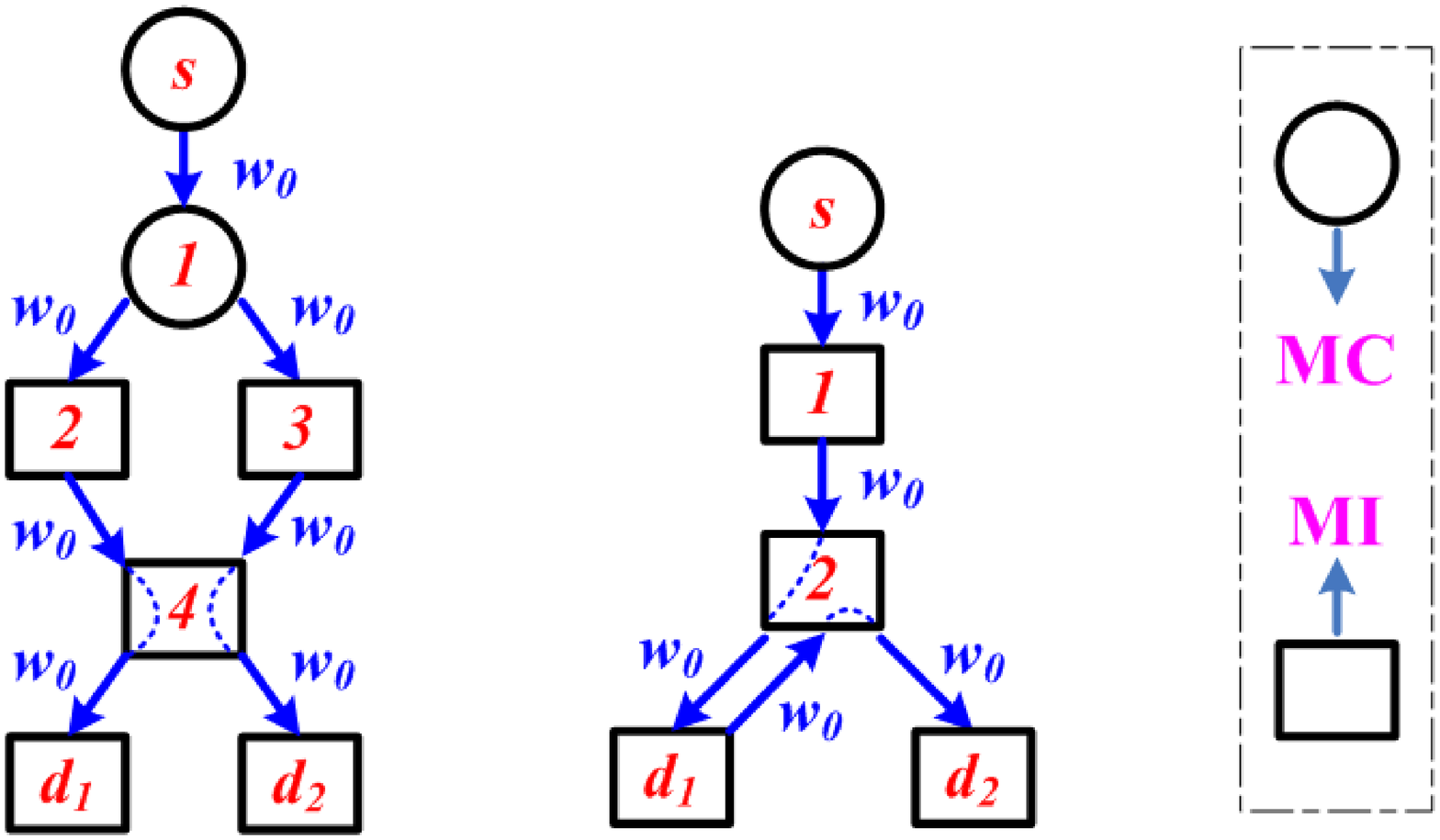}\\
{(a)~~~~~~~~~~~~~~~~~~~~~~~(b)~~~~~~~~~~~~~~~~ }\\
\caption{Two typical light-hierarchies with Cross Pair Switching}
\label{fig: type-LH}
\end{figure}

\subsection{Light-hierarchy Structure}
\label{subsec: Light-hierarchy Structure}
A light-hierarchy is a set of consecutive and directed fiber links occupying the same wavelength, which is rooted from the source and terminated at the destinations. Different from a light-tree, light-hierarchy is free of the repetition of nodes while it forbids the duplicate of the same link. It can be expressed as an enumeration of nodes and links, for instance the light-hierarchy (LH) in Fig.~\ref{fig: type-LH}(a) can be given by
$LH=\big(s(l_{s1},1( l_{12},2( l_{24},4( l_{4d_1}, d_1)), l_{13}, 3( l_{34},4( l_{4d_2}, d_2))))\big)$.

Generally a light-hierarchy has the following characters:
(a) Each link is directed and can be used only once.
(b) Each link has one and only one predecessor link, except that the links coming from source have no predecessor link.
(c) Cycles are permitted.
(d) Only one wavelength is occupied over all the links.
(e) Between each pair of nodes in a light-hierarchy, there are at most two links. Two links are only permitted in condition that they are used for opposite direction communications.
(f) The number of input and output links of a node varies according to its splitting capacity. For a non-terminal MI node, multiple incoming links are allowed. However, each incoming link should correspond to distinct outgoing link. Hence, the number of input links of a non-terminal MI node should be equal to that of its output links. Besides, an MC node should have one and only one input link while it can have as many output links as it can.

Two typical light-hierarchies with Cross Pair Switching are demonstrated in Fig.~\ref{fig: type-LH}, where the circle denotes an MC node while the rectangle stands for an MI node. Source node $s$ multicast messages to destinations $d_1$ and $d_2$. In Fig.~\ref{fig: type-LH}(a), the light signal emitted by $s$ is split into 2 copies by MC node 1, then these two copies enter two different incoming ports of MI node 4 and are switched to destinations $d_1$ and $d_2$ respectively. This kind of Cross Pair Switching benefits from the high degree of MI node 4 (with a degree of at least 4). While in Fig.~\ref{fig: type-LH}(b) the light signal first goes out from MI node 2 to destination $d_1$ and returns back to it after a round tip in the edge between nodes 2 and $d_{1}$, i.e. $link(2,d_{1})$ and $link(d_{1},2)$. The light signal is then forwarded to destination $d_2$. This Cross Pair Switching is based on the simultaneous usage of two oppositely directional fiber links. However, Cross Pair Switching is not always necessary, and thus the light-tree structure can also be viewed as a special light-hierarchy without Cross Pair Switching.

\begin{theorem}
To minimize the wavelength channel cost for a multicast session under sparse splitting constraint, the light-tree structure is not optimal.
\end{theorem}

\begin{figure}
        \centering
        \includegraphics[width=.4\textwidth]{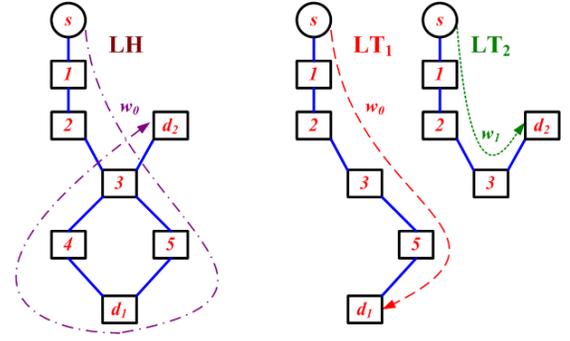}
        \caption{(a) An example network topology and the light-hierarchy; (b) The optimal light-trees for $ms\big(s, (d_{1}, d_{2})\big)$}
        \label{fig: example2}
\end{figure}

\begin{IEEEproof}
Consider the topology in Fig.~\ref{fig: example2}(a) (drawn with solid line), a multicast session $ms\big(s, (d_{1}, d_{2})\big)$ arrives. To implement this session, two light-trees should be constructed. The optimal light-forest solution (i.e., a set of light-trees) is shown in Fig.~\ref{fig: example2}(b): $LT_{1}=\{s-1-2-3-5(or 4)- d_{1}\}$ and $LT_{2}=\{s-1-2-3-d_{2}\}$. The total cost of the optimal light-trees is 9. However, by using the Cross Pair Switching capability of MI node 3, a light-hierarchy (plot in dash-dot line in Fig.~\ref{fig: example2}(a)) could be found out: $LH=\{s-1-2-3-5-d_{1}-4-3-d_{2}\}$. 
The total cost of this light-hierarchy is 8, which is one smaller than that of the optimal light-trees built. We can also see that as the distance between the source $s$ and node 3 becomes bigger, more total cost will be saved.
Hence, in this case, the light-hierarchy structure outperforms the light-tree structure.
\end{IEEEproof}

\begin{theorem}\label{thrm01}
The cost optimal multicast routing structure for sparse splitting WDM networks is a set of light-hierarchies (at least one).
\end{theorem}

\begin{IEEEproof}
It is trivial that a multicast session may be established on several wavelengths. Next we prove that the projection of the cost optimal structure on each used wavelength is a light-hierarchy. Consecutive links in a light-hierarchy assure its connectivity, and direction of links guarantees that the signal could be reached from the source to destinations. Hence, to prove its optimality, it is sufficient to prove that each link should has only one predecessor link in the cost optimal structure. Suppose each link in the optimal structure has two predecessor links. If one predecessor link is removed, then the connectivity and communication can still be guarantied. Thus, it is not cost optimal and this contradicts with the assumption. So, $Theorem$~\ref{thrm01} follows.
\end{IEEEproof}

\section{ILP Formulation of Light-Hierarchy}
\label{sec: ILP Formulation of Light-Hierarchy}
With the help of Cross Pair Switching of MI nodes, the splitting constraint could be relaxed to some extent. Consequently more destinations may be served in one light-hierarchy. This is why the light-hierarchy structure can achieve the optimal cost. In this section, the integer linear programming (ILP) method is applied to search the cost optimal light-hierarchy solutions.\\
\textbf{Network Parameters}:
\begin{flushleft}
\begin{tabular}{ll}
$G$                 &: The graph of the network topology.\\
$V$                 &: The node set of the network $G$.\\
$E$                 &: The edge set of the network $G$.\\
$W$                 &: The set of wavelengths supported per fiber.\\
$\Delta$            &: An integer big enough such that $\Delta>|W|$.\\
$\lambda$           &: A wavelength, $\lambda \in W$.\\
$In(m)$             &: The set of nodes which has a outgoing link\\
                    &  leading to node $m$.\\
$Out(m)$            &: The set of nodes which can be reached from\\
                    &  node $m$.\\
$Deg(m)$            &: The in (or out ) degree of node $m$ in $G$, \\
                    &  where $Deg^{-}(m) = Deg^{+}(m)=Deg(m)$.\\
$link(m,n)$         &: The directed link from node $m$ to node $n$.\\
$e(m,n)$            &: The edge connecting nodes $m$ and $n$ in $G$.\\
                    &  It consists of $link(m,n)$ and $link(n,m)$.\\
$c_{m,n}$           &: The cost of the link from node $m$ to node $n$.\\
$MC\_SET$          &: The set of MC nodes in $G$.\\
$MI\_SET$          &: The set of MI nodes in $G$.\\
\end{tabular}
\end{flushleft}

\textbf{ILP Variables}:
\begin{flushleft}
\begin{tabular}{ll}
$L_{m,n}(\lambda)$  &: Binary variable. Equals to 1 if multicast request\\
                    &   $ms(s,D)$ uses wavelength $\lambda$ on $link(m,n)$,\\
                    &  equals to 0 otherwise.\\
$F_{m,n}(\lambda)$  &: Commodity flow. Denotes the number of \\
                    & destinations served by $link(m,n)$ on $\lambda$.\\
$w(\lambda)$        &: Binary variable. Equals to 1 if $\lambda$ is by the  \\
                    & light-hierarchies, equals to 0 otherwise.\\
\end{tabular}
\end{flushleft}

\subsection{ILP Formulation}
\label{subsec: ILP Formulations}


The principle objective of our problem is minimizing the total cost for a multicast session $ms(s,D)$. Secondly, among the cost optimal light-hierarchy solutions, the one requiring the fewest wavelengths is favorable. To achieve this, a big enough integer $\Delta$ is introduced, which is superior to the number of wavelengths supported per fiber link, i.e. $\Delta > |W|$. Hence the general objective function can be expressed as follows:
\begin{equation}
Minimize: \Delta \times \sum_{\lambda \in W}\sum_{m \in V}\sum_{n \in In(m)}{ c_{n,m}\cdot L_{n,m}(\lambda)} + \sum_{\lambda \in W}{w(\lambda)}
\end{equation}
This objective function is subject to a set of constraints, which are listed below:

\subsubsection{Light-Hierarchy Structure Constraints}
~\\Source Constraint:
\begin{eqnarray}
\label{equation: LH-sc-1}
\sum_{\lambda \in W}\sum_{n \in In(s)}{L_{n,s}(\lambda)} = 0\\
\label{equation: LH-sc-2}
1 \leq \sum_{\lambda \in W}\sum_{n \in Out(s)}{L_{s,n}(\lambda)} \leq |D|
\end{eqnarray}

Constraints (\ref{equation: LH-sc-1}) and (\ref{equation: LH-sc-2}) ensure that the light-hierarchies for multicast session $ms(s,D)$ are rooted at the source node $s$. The source $s$ must not have any input link in a light-hierarchy, but must at least one output link on some wavelength and the total number of links going out from $s$ should not go beyond the number of sink nodes, i.e., $|D|$.

Destination Constraint:
\begin{eqnarray}
\label{equation: LH-dc-1}
1 \leq \sum_{\lambda \in W}\sum_{n \in In(d)}{L_{n,d}(\lambda)} \leq |D|-1, & \forall d \in D
\end{eqnarray}

Constraint (\ref{equation: LH-dc-1}) guarantees that each destination node should be spanned in at least one but at most $|D|-1$ light-hierarchies.
%


MC node Constraint:
\begin{eqnarray}
\label{equation: LH-MC-Input}
\sum_{n \in In(m)}{L_{n,m}(\lambda)} \leq 1,\\
\label{euqation: LH-MC-Output}
\sum_{n \in Out(m)}{L_{m,n}(\lambda)} \leq Deg(m) \times\sum_{n \in In(m)}{L_{n,m}(\lambda)}, \\
\nonumber \forall \lambda \in W, \forall m \in MC\_SET \textrm{ and } m \neq s
\end{eqnarray}

Constraint (\ref{equation: LH-MC-Input}) makes sure that each MC node has only one input link. This constraint together with constraint (\ref{euqation: LH-MC-Output}) also indicates that if and only if an MC node $m$ is spanned in a light-hierarchie, then the number of outgoing links of $m$ is between 1 and $Deg(m)$. Otherwise, the number of outgoing link of $m$ must be 0.

MI node Constraint:
\begin{eqnarray}
\label{equation: LH-MI-Output}
\sum_{n \in Out(m)}{L_{m,n}(\lambda)} \leq \sum_{n \in In(m)}{L_{n,m}(\lambda)},\\
\nonumber \forall \lambda \in W, \forall m \in MI\_SET \textrm{ and } m \neq s
\end{eqnarray}
Since the number of input links is not restricted, MI nodes are enabled to make the Cross Pair Switching. According to equations (\ref{equation: LH-MI-Output}) and (\ref{equation: LH-Leaf}), MI nodes are allowed to branch in condition that the number of incoming branches equals the number of outgoing branches if they are non-member nodes. Nevertheless, the MI destination nodes may not have any outgoing branches.

Leaf Node Constraint:
\begin{eqnarray}
\label{equation: LH-Leaf}
\sum_{n \in Out(m)}{L_{m,n}(\lambda)} \geq \sum_{n \in In(m)}{L_{n,m}(\lambda)}\\
\nonumber \forall \lambda \in W, \forall m \in V \textrm{ and } m \not \in D
\end{eqnarray}
Constraint (\ref{equation: LH-Leaf}) ensures that only the destination nodes can be leaf nodes in the light-hierarchies while the non-member nodes can not.

Relationship between $L_{m,n}(\lambda)$ and $w(\lambda)$:
\begin{eqnarray}
\label{equation: CC-Relation3}
w(\lambda) \geq L_{m,n}(\lambda), \forall m,n \in V, \forall \lambda \in W\\
\label{equation: CC-Relation4}
w(\lambda) \leq \sum_{\forall m \in V}\sum_{\forall n \in V}L_{m,n}(\lambda), \forall \lambda \in W
\end{eqnarray}

\begin{figure}
\centering
\includegraphics[width=3.5in]{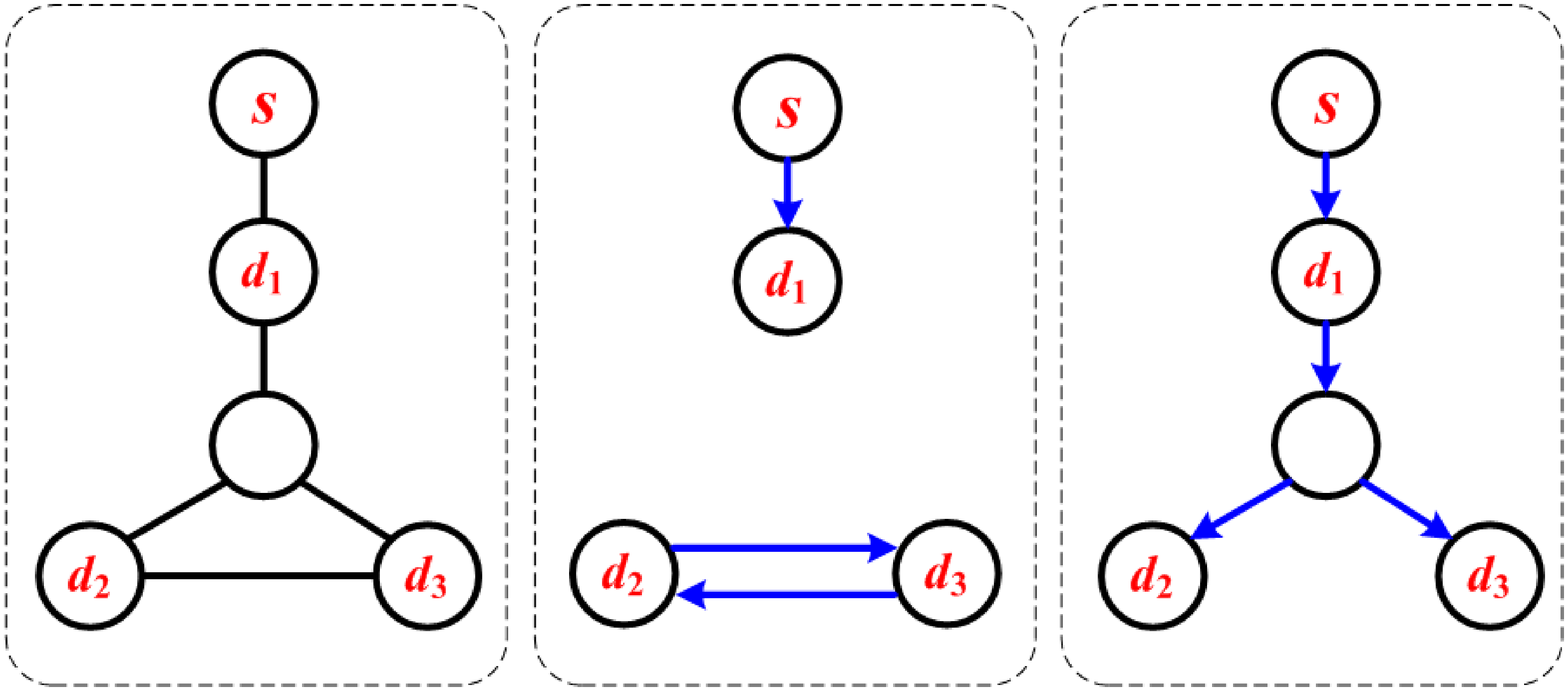}\\
\mbox{(a)~~~~~~~~~~~~~~~~~~~~~(b)~~~~~~~~~~~~~~~~~~~~~(c)}
\caption{(a) An example network topology; (b) The false result; (c) The optimal result.}
\label{fig: Loop}
\end{figure}

The light-hierarchy structure constraints above are not sufficient to guarantee that the resultant light-hierarchy be connected. For instance, $ms\big(s, (d_{1}-d_{3})\big)$ is required in topology Fig.~\ref{fig: Loop}(a). By just applying the light-hierarchy structure constraints, the optimal solution is shown in Fig.~\ref{fig: Loop}(b): $L_{s,d_{1}}(\lambda_{1})=1$, $L_{d_{2},d_{3}}(\lambda_{1}) =1$, $L_{d_{3},d_{2}}(\lambda_{1}) =1$ while the other variables $L_{m,n}(\lambda)=0$. Unfortunately the result is incorrect, although this result complies the light-hierarchy constraints above. This is because destinations $d_{2}$ and $d_{3}$ are not reachable from the source node 1. In \cite{yOliver2005}, a commodity flow method was proposed to search the loss-balanced light-tree. Here, we apply this method to create supplementary formulations to restrain the variables $L_{n,m}(\lambda)$ in order that the continuity and the connectivity of the resultant light-hierarchy could be guaranteed.

\subsubsection{Connectivity Constraints}
~\\To establish a multicast session, several light-hierarchies may be required. And the same destination may be spanned by several light-hierarchies. However, a destination can only be served in one light-hierarchy to consume the light signal (i.e. receive the multicast messages), while it is spanned in the other light-hierarchies to uniquely forward the light signal to the successor node.

Source node:
\begin{eqnarray}
\label{equation: CC-sc}
\sum_{\lambda \in W}\sum_{n\in Out(s)}F_{s,n}(\lambda) = |D|
\end{eqnarray}
Constraint (\ref{equation: CC-sc}) indicates that the sum of the commodity flow emitted by the source should be equal to $|D|$ the number of destinations in the multicast session.

Destination nodes:
\begin{eqnarray}
\label{equation: CC-dn-1}
\sum_{\lambda \in W}\sum_{n\in In(d)}F_{n,d}(\lambda) = \sum_{\lambda \in W}\sum_{n\in Out(d)}F_{d,n}(\lambda) + 1, \forall d \in D
\end{eqnarray}
\begin{eqnarray}
\label{equation: CC-dn-2}
\nonumber \sum_{n\in In(d)}F_{n,d}(\lambda)-1 \leq \sum_{n\in Out(d)}F_{d,n}(\lambda) \leq \sum_{n\in In(d)}F_{n,d}(\lambda), \\
\forall d \in D, \forall \lambda \in W
\end{eqnarray}
Equations (\ref{equation: CC-dn-1}) and (\ref{equation: CC-dn-2}) ensure that each destination node should consume totally one and only one flow in all the light-hierarchies. This constraint also guarantees that each destination is reachable from the source $s$.

Non-Member nodes:
\begin{eqnarray}
\label{equation: CC-noMem}
\sum_{n\in In(m)}F_{n,m}(\lambda) = \sum_{n\in Out(m)}F_{m,n}(\lambda), \\
\nonumber \forall m \in V\setminus(s\cup D),\forall \lambda \in W
\end{eqnarray}
Equation (\ref{equation: CC-noMem}) guarantees that the flow does not drop after passing a non-member node.

Relationship between $L_{m,n}(\lambda)$ and $F_{m,n}(\lambda)$:
\begin{eqnarray}
\label{equation: CC-Relation1}
F_{m,n}(\lambda) \geq L_{m,n}(\lambda), \forall m,n \in V, \forall \lambda \in W\\
\label{equation: CC-Relation2}
F_{m,n}(\lambda) \leq |D| \times L_{m,n}(\lambda), \forall m,n \in V, \forall \lambda \in W
\end{eqnarray}
Equations (\ref{equation: CC-Relation1}) and (\ref{equation: CC-Relation2}) show that a link should carry non-zero flow if it is used in a light-hierarchy, and the value of this flow should not beyond the total flow emitted by the source node.

\section{Simulation and Performance Evaluation}
\label{sec: Simulation and Performance Evaluation}
\subsection{Simulation Model}
\label{subsec: Simulation Model}
\begin{figure} [!t]
    \centering
    \includegraphics[width=2.5in]{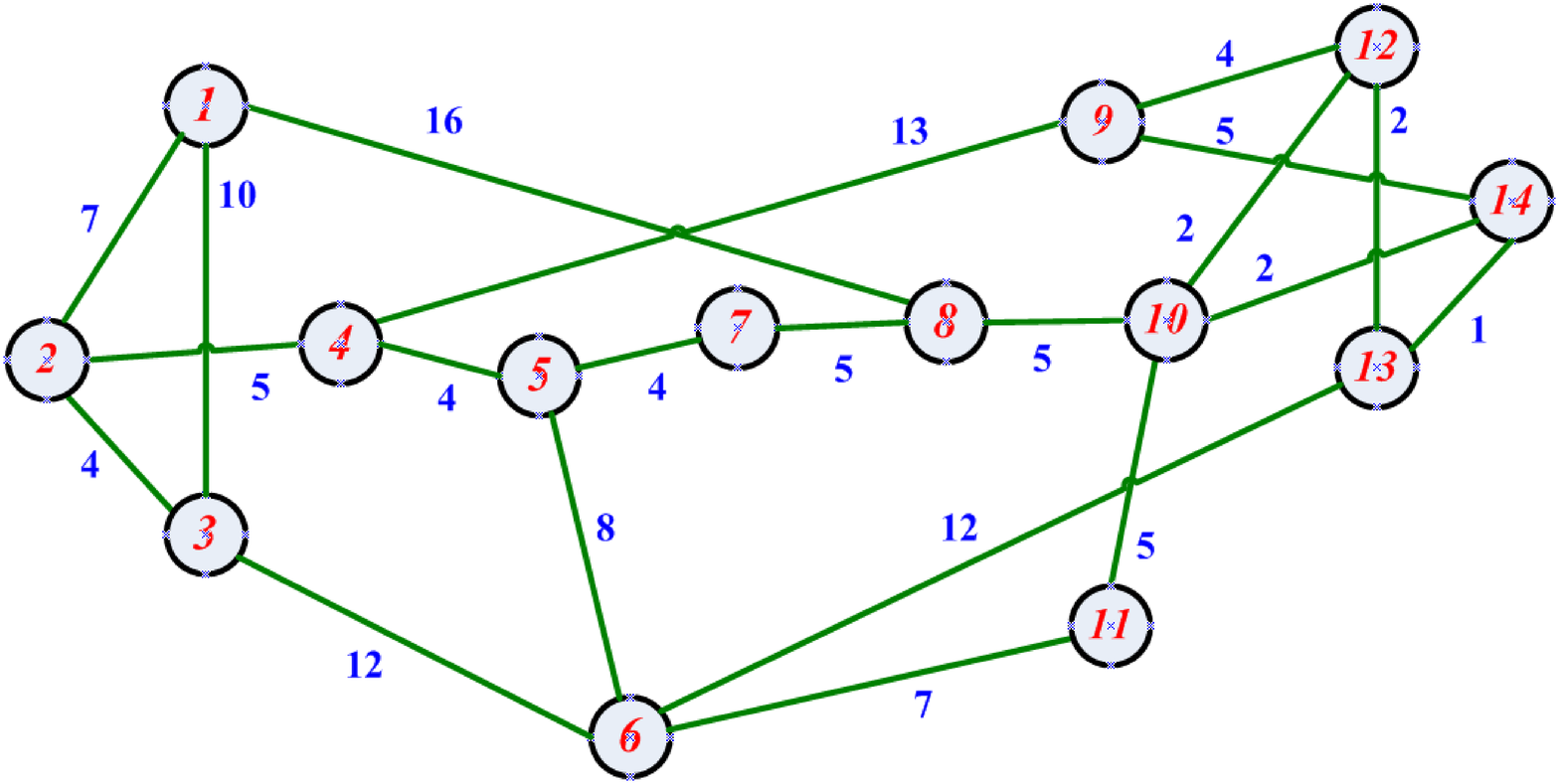}
    \caption{NSF Network}
    \label{fig_NSF}
  \end{figure}

\begin{figure} [!t]
    \centering
    \includegraphics[width=2in]{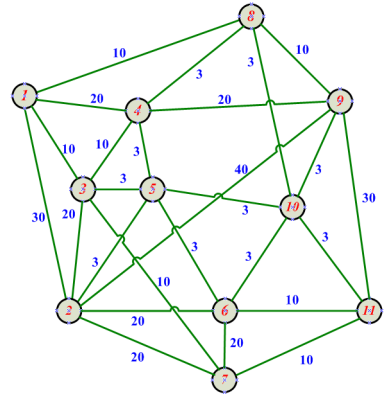}
    \caption{European Cost-239 Network}
    \label{fig_cost239}
  \end{figure}

In order to demonstrate the advantage of the proposed light-hierarchy structure, simulation is conducted to compare it with the light-tree structures. ILP formulations are implemented in C++ with Cplex package by using the 14-nodes NSF network in Fig.~\ref{fig_NSF} and 11-nodes European Cost-239 network in Fig.~\ref{fig_cost239}. Given a group size $|D|$, 100 random multicast sessions are generated. The membership of each multicast session follows a uniform distribution in the topology. Then, ILP formulations are executed to search the optimal light-trees and the light-hierarchies with the minimum cost for each multicast session.

\subsection{Light-Hierarchy versus Light-tree}
\label{subsubsec: Light-Hierarchy versus Light-tree}
To show the applicability of the light-hierarchy structure, the performances of light-hierarchy (LH) and its counterpart light-tree (LT) are simulated in both NSF network and Cost-239 network. The following metrics are taken into account:
(1) Total cost consumed for the establishment of 100 sessions, as well as the cost saving percentage of light-hierarchy (LH) structure compared to light-tree structure (LT).
(2) The number of wavelengths required for 100 sessions.
(3) R(CPS), the number of light-hierarchies employing the Cross Pair Switching.

The numerical results are presented in Tables~\ref{table_NSF} and \ref{table_Cost239}. Two cases are considered. Case A stands for no splitter while Case B stands for sparse splitting. Based on the simulation results, it is observed that:
 (a) The proposed light-hierarchy structure always achieve much lower total cost than the traditional light-tree structure. The cost can be saved up to 3.61\% by 57 light-hierarchies with $CPS$ in NSF network, while up to 6.27\% by 81 light-hierarchies with $CPS$ in Cost-239 network. Therefore, light-tree structure is not optimal from the point view of cost, but the light-hierarchy structure can be better. (b) In general, the absolute cost reduction by the light-hierarchy structure depends on the number of Cross Pair Switching used, i.e. R(CPS). This is because that with the help of Cross Pair Switching of MI nodes a destination may connect to the light-hierarchy with less cost by using cycles. (c) Fewer wavelengths on average are required for establishing 100 multicast sessions, when the light-hierarchy method is adopted.

 All of these advantages benefit from the proposed Cross Pair Switching capability of MI nodes. The light-tree structure restraints that each node should have only one input link, while the light-hierarchy structure accepts cycles rather than complying this old rule. Since the new light-hierarchy structure overcomes the inherent shortcoming of the tree structure, it is able to make the most of Cross Pair Switching by employing the incoming and outgoing link pairs of MI nodes. Therefore, more destination nodes can be served in one light-hierarchy than a light-tree, and thus fewer wavelengths is required by each session.  With the help of the light-hierarchy structure, a destination node is more likely to connect to the nearest node (even if it is an MI node) in the light-hierarchy while it may have to lead a long way to the source node on another wavelength in order not to violate the light-tree structure. As the light-tree is a special type of light-hierarchy, the optimal light-hierarchy solution at least has the same cost as the light-tree solution in the worst cases. Once useful Cross Pair Switching node is found, the total cost is decreased. More Cross Pair Switching is used, more cost will be saved. This explains the third observation.

%

\begin{table}[!t]
\caption{Performance comparison in USA NSF Network.}
\label{table_NSF}
\centering
\begin{tabular}{|c|c||c|c|c|c|c|}
\hline
\multicolumn{7}{|c|}{Case A: No splitters. }\\
\hline
\hline
$Size$ & \multicolumn{3}{c|}{Total Cost} & \multicolumn{2}{c|}{Wavelengths}& LH \\
\hline
$|D|$ & LH & LT & $\searrow$ &LH & LT & R(CPS) \\
\hline
\hline
2 & 2059 &  2079 & 0.96\% & 103  & 106   & 10/100\\
\hline
6 & 4096 &  4247 & 3.56\% & 107  & 114   & 35/100\\
\hline
9 & 5025 &  5213 & 3.61\% & 115  & 147  & 57/100\\
\hline
13 & 6237 &  6330  & 1.47\% & 121  & 156  & 67/100\\
\hline
\hline
\multicolumn{7}{|c|}{Case B: nodes 5 and 8 are splitters (MC nodes) }\\
\hline
\hline
$Size$ & \multicolumn{3}{c|}{Total Cost} & \multicolumn{2}{c|}{Wavelengths}& LH \\
\hline
$|D|$ & LH & LT & $\searrow$ & LH & LT & R(CPS) \\
\hline
\hline
2 & 2055 &  2075 & 0.96\% &  103  & 106   & 11/100\\
\hline
6 & 4017 &  4080 & 1.54\% &  105  & 108   & 32/100\\
\hline
9 & 4898 &  4984 & 1.73\% &  105  & 112  & 36/100\\
\hline
13 & 6035 &  6035  & 0\% & 106  & 111  & 5/100\\
\hline
\end{tabular}
\end{table}

\begin{table}[!t]
\caption{Performance comparison in European Cost-239 Network.}
\label{table_Cost239}
\centering
\begin{tabular}{|c|c||c|c|c|c|c|}
\hline
\multicolumn{7}{|c|}{Case A: No splitters. }\\
\hline
\hline
$Size$ & \multicolumn{3}{c|}{Total Cost} & \multicolumn{2}{c|}{Wavelengths}& LH \\
\hline
$|D|$ & LH & LT & $\searrow$ &LH & LT & R(CPS) \\
\hline
\hline
2 & 1341 &  1364 & 1.68\% &  100  & 108   & 22/100\\
\hline
5 & 2691 &  2871 & 6.27\% &  104  & 183   & 81/100\\
\hline
7 & 3580 &  3747 & 4.46\% & 100  & 223  & 93/100\\
\hline
10 & 5204 &  5336 & 2.47\%  & 120  & 272  & 100/100\\
\hline
\hline
\multicolumn{7}{|c|}{Case B: nodes 3 and 9 are splitters (MC nodes) }\\
\hline
\hline
$Size$ & \multicolumn{3}{c|}{Total Cost} & \multicolumn{2}{c|}{Wavelengths}& LH \\
\hline
$|D|$ & LH & LT & $\searrow$ & LH & LT & R(CPS) \\
\hline
\hline
2 & 1329 &  1344 & 1.12\% &  100  & 108   & 16/100\\
\hline
5 & 2685 &  2863 & 6.22\% &  102  & 183   & 82/100\\
\hline
7 & 3580 &  3747 & 4.46\% & 100  & 223  & 93/100\\
\hline
10 & 5204 &  5280  & 1.44\% & 100  & 272  & 100/100\\
\hline
\end{tabular}
\end{table}

\section{Conclusion}
\label{sec: Conclusion}
Instead of the traditional light-tree structure, a new all-optical multicast structure called light-hierarchy is proposed to improve the quality of multicast routing in sparse splitting WDM networks. A light-hierarchy is a set of consecutive and directed fiber links occupying the same wavelength, which is rooted from the source and terminated at the destinations. Different from a light-tree, the light-hierarchy structure accepts the cycles introduced by the Cross Pair Switching capability of MI nodes, which enables an MI node to serve several destination nodes on the same wavelength through its different input and output pairs. Light-hierarchy structure overcomes the inherent drawback of the traditional light-tree structure, so that the splitting constraint is relaxed to some extent. This is why it outperforms the light-tree in term of cost. We proved that the optimal multicast structure for minimizing the wavelength channel cost is not a set of light-trees, but light-hierarchies. ILP formulations are developed and implemented to compute the optimal light-hierarchies. Numerical results verified that the light-hierarchy structure is the cost optimal solution for all-optical multicast routing with sparse splitting constraint.


%
%



%

\end{document}